\documentclass[aps,english,prb,twocolumn,showpacs,superscriptaddress]{revtex4}
\usepackage{graphicx}
\usepackage[left]{lineno}
\usepackage{gensymb}

\begin{document}

\title{Symmetry induced stability in alkali doped calcium-silicate-hydrate}

\author{V. Ongun {\"O}z{\c{c}}elik}\email{ongun@princeton.edu}
\affiliation{Andlinger Center for Energy and the Environment, Princeton University, New Jersey 08544, USA}
\affiliation{Civil and Environmental Engineering, Princeton University, New Jersey 08544, USA}

\author{Nishant Garg} 
\affiliation{Andlinger Center for Energy and the Environment, Princeton University, New Jersey 08544, USA}
\affiliation{Civil and Environmental Engineering, Princeton University, New Jersey 08544, USA}

\author{Claire E. White}\email{whitece@princeton.edu}
\affiliation{Andlinger Center for Energy and the Environment, Princeton University, New Jersey 08544, USA}
\affiliation{Civil and Environmental Engineering, Princeton University, New Jersey 08544, USA}

\begin{abstract}
CO$_2$ emissions originating from the construction industry have a significant impact on global warming where the production of ordinary Portland cement clinker is responsible for approximately 8\% of all human-made CO$_2$. Alkali doped calcium-silicate-hydrate (C-S-H) is a critical silicate material since the use of blended cements and alkali-activated materials in construction industry can substantially reduce human-made CO$_2$ emissions. However, the effect of alkali doping (Na and K) on the long-term stability and associated durability of C-S-H remains an open question. Here, using first principles quantum chemistry calculations on the model crystalline phase clinotobermorite, we show that there is a strong interplay between the thermodynamic stability of alkali doped C-S-H and the symmetry of the alkali atoms in the structure. Our results reveal that a symmetrical distribution of alkali atoms leads to a higher stability value such that stable structures with moderate alkali concentrations can be obtained provided that the alkali atoms are allowed to settle into a symmetrical distribution.  We investigate the associated structural mechanisms by calculating the migration barriers of alkali atoms within the material, the electronic charge distribution in the material and the variation of basal spacing  by using both computational methods and X-ray diffraction analysis. 
\end{abstract}

\maketitle

Calcium-silicate-hydrate (C-S-H) is a versatile silicate material which is mainly responsible for the strength and durability of cement paste.\cite{allen2007composition} Although widely studied,\cite{faucon1999aluminum, pustovgar2016understanding, andersen2003incorporation,rieger2014formation,l2016alkali,l2016influence, miller2018impacts, kumar2017atomic,goracci2017dynamics,krautwurst2018two} experimental techniques fail to reveal the accurate atomic structure of C-S-H due to its highly amorphous nature, varying range of Ca/Si ratios and difficulties in separating it from other phases. On the theoretical front, several structural models have been suggested \cite{skinner2010nanostructure, myers2013generalized, richardson1999nature, qomi2014combinatorial, puertas2011model,ozccelik2016nanoscale} for C-S-H, however each of these models require additional verification and their thermodynamic stabilities should be separately assessed under specific conditions (such as effects of elevated temperatures, inclusion of impurities, and exposure to different chemical environments). Among these factors, the inclusion of impurities in the models (such as alkalis and aluminum) further increases the complexity of the proposed atomic arrangements.\cite{white2015intrinsic, ozccelik2016nanoscale, arayro2018thermodynamics}

Augmenting C-S-H by the inclusion of alkali atoms (such as Na or K) in the pure structure is a critically important example where the chemical composition of C-S-H is modified on purpose. Alkali doping of C-S-H has implications in various applications ranging from obtaining new blended systems to synthesizing high-Ca alkali-activated materials (AAMs). AAMs are particularly important since they have been proposed as an alternative to ordinary Portland cement (OPC) which is responsible for approximately 8\% of human-made CO$_2$ emissions. \cite{scrivener2008innovation,van2016long,monteiro2017towards} By using AAMs it is possible to reduce the CO$_2$ emissions associated with the industry by up to 80\% as compared to using OPC. \cite{duxson2007role,mclellan2011costs,myers2013generalized} However, it is imperative to understand the impact of alkalis on the stability of pure and aluminum-substituted C-S-H since this behavior will influence the long-term durability (i.e., performance) associated with use of AAM concrete in industry. Furthermore, the use of alkali-containing supplementary cementitious materials in OPC systems is becoming more common (e.g., ground recycled glass), and therefore increasing attention is being paid to the alkali-silica reaction in concrete. However, if the alkalis could be encapsulated in C-S-H then they would not cause the potentially catastrophic alkali-silica reaction. Therefore, an accurate understanding of the fundamental mechanisms that control the long term durability of these new alkali-containing systems will have a significant impact on reducing human-made CO$_2$ emissions. In a recent letter,\cite{ozccelik2016nanoscale} using first-principles quantum chemical calculations, we revealed the existence of a charge balancing mechanism at the atomistic level that contributes to the macroscopically measured properties of alkali-activated cement-based gels. These computational models showed that the long term phase stability and nano-mechanical strength of Ca-rich AAMs are in the same range of C-S-H-based OPC systems, provided that proper charge balancing takes place during their synthesis.

\begin{figure*}
\includegraphics[width=12cm]{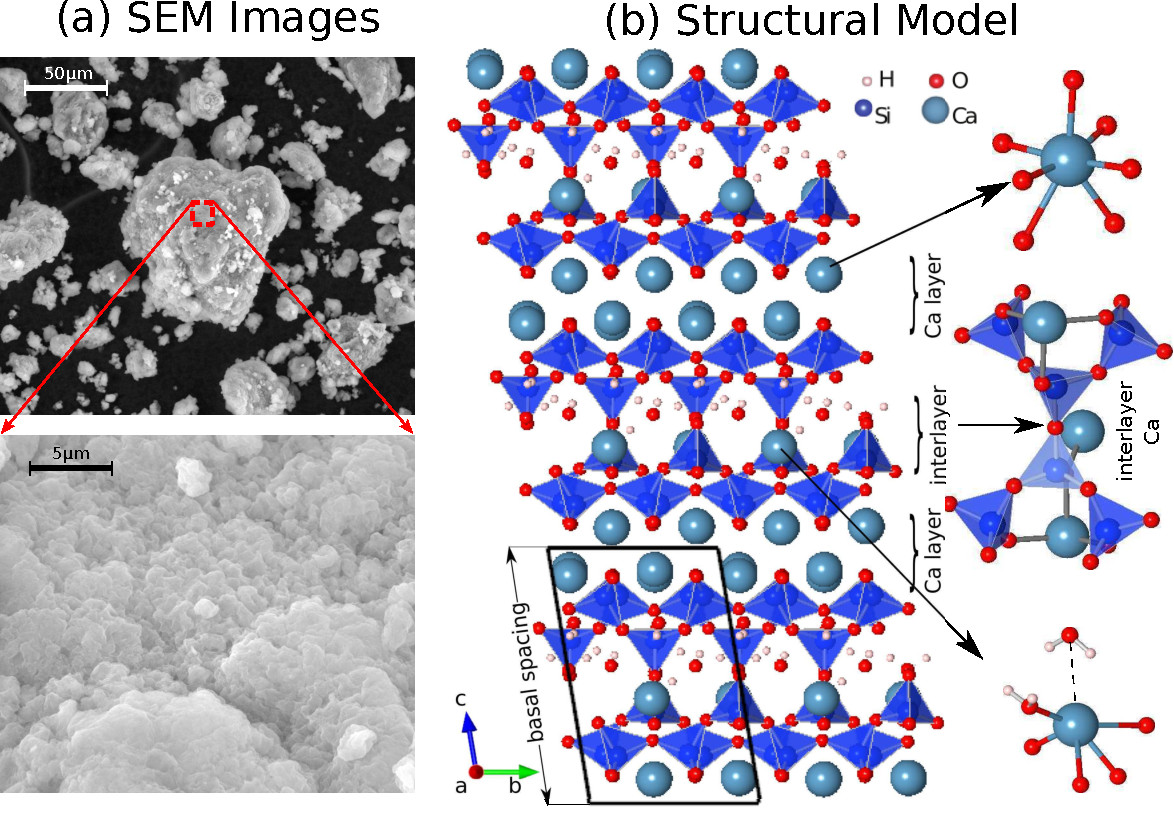}
\caption{(a) Scanning electron microscopy (SEM) images of  the C-S-H gel with a Ca/Si=1.0. (b) Optimized structure of 11 {\AA} clinotobermorite with lattice constants a=11.27 {\AA}, b=7.34 {\AA} and c=11.47 {\AA}. The building blocks of the structure are separately shown on the right panel. In the ball-and-stick model Ca, Si, O and H atoms are shown with light blue, dark blue, red and pink spheres, respectively.}
\label{fig1}
\end{figure*}

Here, using quantum chemical calculations based on density functional theory (DFT)  and  X-ray diffraction (XRD) analysis we reveal that the concentration and geometrical distribution  of the alkali atoms at the atomistic length scale determines the macroscopic properties of the overall C-S-H structure such as its thermodynamic stability and basal spacing. Briefly, we show that there is a geometry dependent interplay between the level of alkali doping and the stability of the C-S-H gel, such that the ground state total energy of the structure decreases as the distribution of alkali atoms becomes more symmetrical. By computing the migration barriers of the alkali atoms in the interlayer region we show that it is possible for the alkali atoms in the C-S-H structure to evolve towards a symmetrical distribution in the interlayer region provided that sufficient time is allowed for the structure to settle. We also show that the basal spacing, which is one of the characteristic properties of C-S-H,  is also closely related with the symmetry of the alkali atoms as well as their concentration.

\begin{figure*}
\includegraphics[width=14cm]{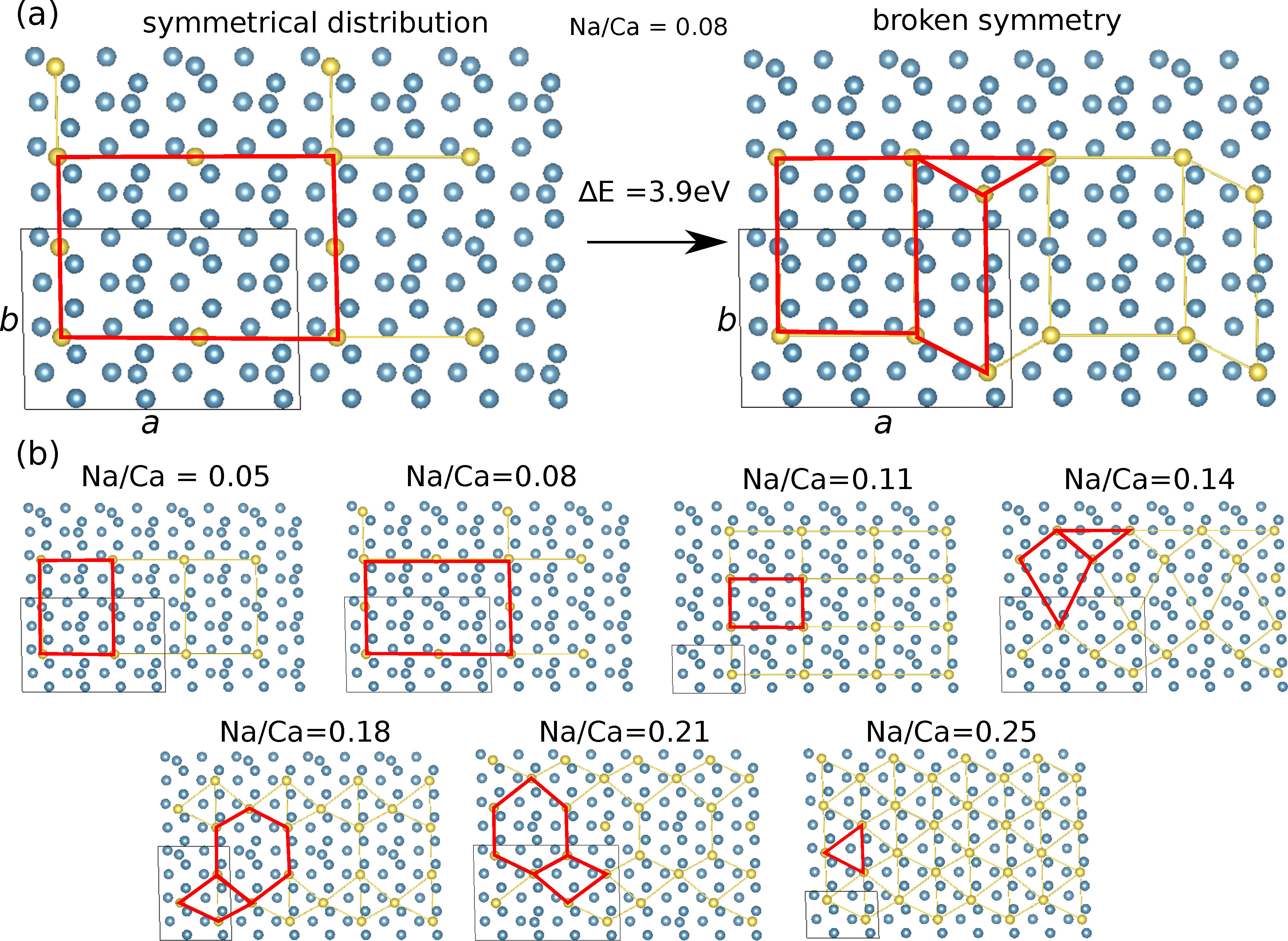}
\caption{Top view of alkali doped clinotobermorite with different Na/Ca ratios. Ca and Na atoms are shown with light blue and yellow spheres, respectively. (a) Two possible distributions of Na atoms in alkali doped clinotobermorite with Na/Ca = 0.08. The structure with symmetrically distributed Na atoms is more favorable than the asymmetric case by 3.9 eV. (b) Optimized ground state distributions of Na in alkali doped clinotobermorite for different Na/Ca ratios. In all structures, the repeating symmetric patterns are highlighted with red lines for better visibility.}
\label{fig2}
\end{figure*}

\subsection*{Morphology and molecular structure}
A total of 45 different C-(N)-A-S-H gels (C-S-H with Al and Na) were synthesized with varying Ca/Si (0.6, 1.0, 1.4) and Al/Si (0.0, 0.05, 0.1) ratios in the presence of NaOH solutions of varying concentrations (0.0, 0.1, 0.5, 1, 5 M). Synchrotron X-ray diffraction was conducted on the bulk gels in order to extract basal spacing data which are discussed and compared with our DFT results in a later section.  The particles in the samples have a wide size distribution ranging from tens to hundreds of microns as shown in the scanning electron microscope (SEM) images in Fig.~\ref{fig1}a. At the nanoscale, the synthetic gels are known to consist of calcium-silicate layers of varying lengths and orientations resulting in gel pore spaces containing confined water.\cite{l2016influence, richardson2014model}

To investigate the fundamental atomistic scale mechanisms underlying the  bulk characteristics of the alkali doped C-S-H, we concentrated on obtaining the optimized atomic structure. A model derived purely from first-principles calculations without any empirical inputs is needed to accurately assess the effect of doping on stability. For this purpose, we focused on the 11{\AA} clinotobermorite since it is thought to be one of the most realistic crystalline models for C-S-H, especially when Q$^3$ Si sites are present as is the case for high-Ca AAMs.  Clinotobermorite, which is a crystalline monoclinic representation of C-S-H, consists of layers of Ca-O polyhedrons with silicate chains on each side of the polyhedrons.\cite{richardson2014model, hoffmann1997clinotobermorite} Additional Ca atoms and water molecules are present between the repeating layers. We optimized the structure by performing self-consistent DFT loops which included calculating the volume and shape of the crystal unit-cell, optimizing the lengths of Si-O and Ca-O bonds, determining the number and orientation of H$_2$O molecules  in the interlayer region with respect to the silicate chains, and finally optimizing the interlayer spacing. After these steps, we obtained the ground state of 11{\AA} clinotobermorite with the chemical formula of Ca$_{10}$Si$_{12}$O$_{32} \cdot$8H$_2$O, having zero internal pressure as presented in Fig.~\ref{fig1}b. This initial structure, which we refer to as the pure structure throughout this paper, is critical for our study since the formation energies of the alkali doped structures have been calculated in reference to this pure structure.

\subsection*{Energetics and effect of symmetry}
For the pure structure, we calculated the cohesive energy per unit-cell defined by $E_{c}=\sum_{i}n_{i}E_{i}- E_{structure}$,\cite{kittel} which is the difference between the sum of energies of free atoms present in the unit-cell of the structure and the total energy of the overall structure in its optimized geometry. Here, $n_i$ is the number of $i$ atoms present in the unit-cell and the ground state energy of each single atom is calculated in its own magnetic state in accordance to Lieb's theorem. \cite{lieb} Consequently, the cohesive energy of the clinotobermorite structure is 5.58 eV per atom, which signifies the average energy gained per atom by constructing a particular clinotobermorite phase from its constituent atoms. This value is larger than the cohesive energy of the well-known  14{\AA} tobermorite, which is found to be 5.15 eV per atom. \cite{ozccelik2016nanoscale} Such a large cohesive energy indicates comparable phase stability with 14{\AA} tobermorite  at ambient conditions.  To assess the effect of alkali doping on stability, we prepared structures with Na/Ca ratios varying between 0-0.25 by substitution of the interlayer Ca atoms with Na atoms until the desired level of doping was achieved. Here we note that in our computational model the Ca/Si ratio varies between 0.83-0.67 whereas the Na/Si ratio varies between 0-0.17. In previous experimental studies, alkali doped C-S-H phases were successfully synthesized with Ca/Si and alkali/Si ratios up to 1.5 and 0.30, respectively\cite{l2016influence} Thus, the alkali doping levels we have investigated fall into the experimentally synthesizable region.

\begin{figure*}
\includegraphics[width=12cm]{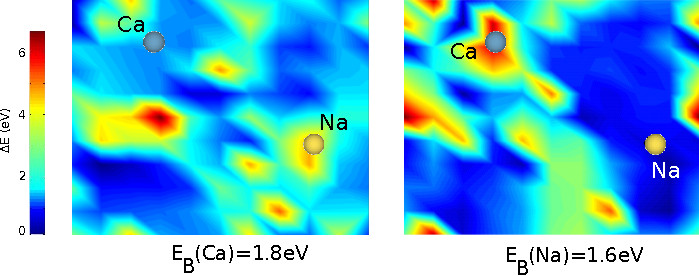}
\caption{Energy landscapes for Ca (left panel) and Na (right panel) atoms in the interlayer region of Na doped clinotobermorite with Na/Ca ratio = 0.25. To migrate in the interlayer region, Na and Ca atoms need to overcome a minimum energy barrier ($E_{B}$) of 1.6 and 1.8 eV, respectively.  In the color map, blue regions represent energetically favorable sites for the migrating atom as compared to the red regions. The dimensions of the landscape correspond to the \textit{ab}-plane of the unit cell.}
\label{fig3}
\end{figure*}

\begin{figure}
\includegraphics[width=8cm]{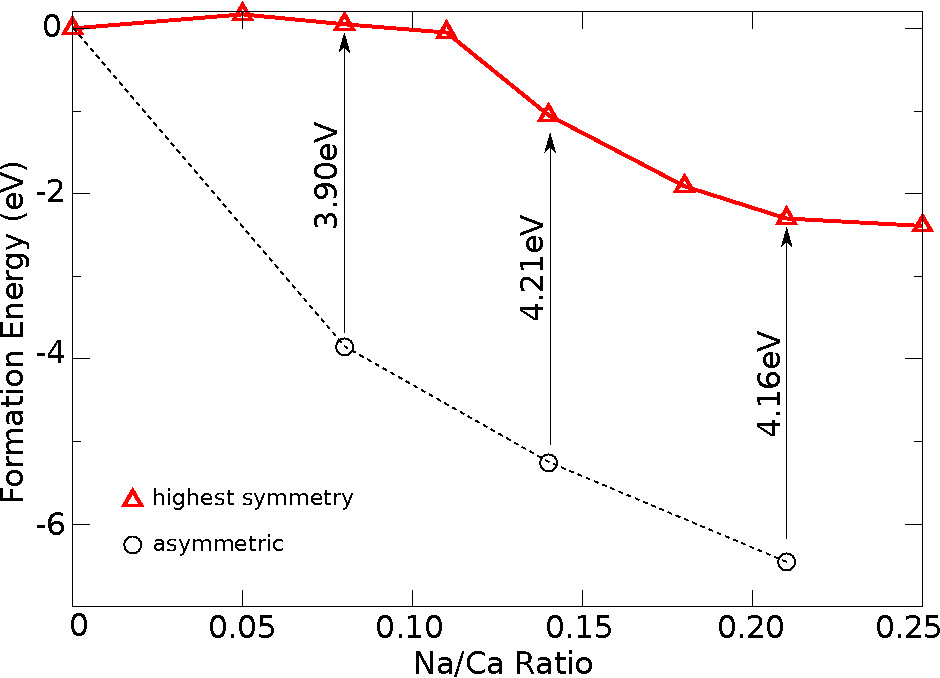}
\caption{Variation of the formation energy of alkali doped clinotobermorite with Na/Ca ratio, where the Na/Ca ratio is calculated using the total number of Na and Ca atoms in each structure. The red and black curves represent structures with highest and lowest symmetries, respectively. Note that, in our notation, negative values of $E_{f}$ imply less stable structures as compared to the pure case.}
\label{fig4}
\end{figure}

\begin{figure*}
\includegraphics[width=10cm]{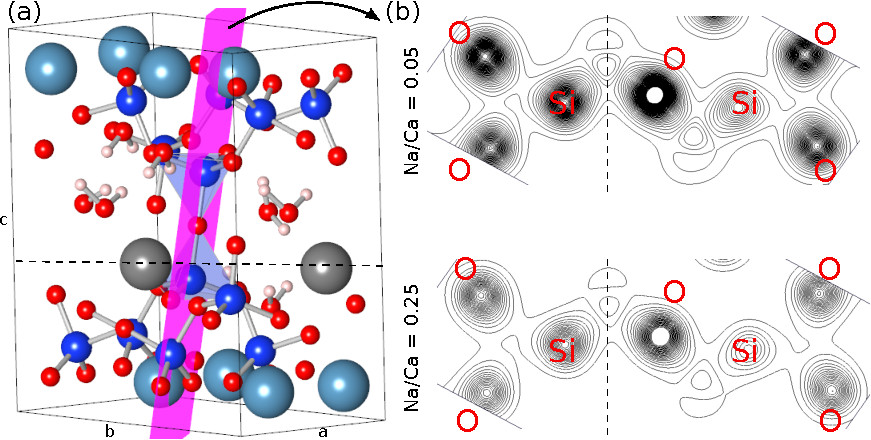}
\caption{Effect of Na concentration on the chemical bonds between the  silicate tetrahedrons in the interlayer region. Total charge density between the connected silicate tetrahedrons are calculated along the (pink) cross section shown in (a). (b) The contour plots of the electron density distribution between the silicate tetrahedrons are shown for Na/Ca=0.05 and Na/Ca=0.25. The dashed lines indicate the position of Ca and Na atoms in the interlayer region.}
\label{fig5}
\end{figure*}

In our model structures, the Na atoms are positioned in the interlayer region since replacing an interlayer Ca atom with an Na is 0.92 eV more favorable than replacing an intralayer Ca atom with Na. Therefore, during the formation of a Ca-rich AAM gel, it is should be expected that Na atoms will tend to be located in the interlayer region rather than penetrating into the intralayer region. However, this doping breaks the charge balance in the vicinity of the interlayer substitution sites due to the ionic nature of Ca and Na in this system, and we are effectively replacing a $^+2$ ion (Ca$^{2+}$) with a $^+1$ ion (Na$^{1+}$). We maintained the charge balance by introducing an external H atom to the interlayer region in the local vicinity of the substitution site. This method was previously shown to be an effective method for proper charge balancing of the alkali doped 14{\AA} tobermorite structure.\cite{ozccelik2016nanoscale} After Na substitution and proper charge balancing, each structure was re-optimized using the same set of parameters that was used for the pure case. Note that our choice of Na as the alkali atom is due to the fact that it is abundant on the earth's crust, but the results presented hereafter should be similar for other $^+1$ alkalis.

Since there are numerous possible substitution sites in the interlayer region, the same level of doping can be achieved with different arrangements of Na atoms depending on which Ca atoms are substituted. For instance, in Fig.~\ref{fig2}a we present two possible arrangements of Na atoms within the interlayer region which both lead to an Na/Ca ratio of 0.08. The difference between the two cases presented in Fig.~\ref{fig2}a is that the structure on the left has a symmetrical distribution of Na atoms in the interlayer region whereas the symmetry is broken for the structure presented on the right panel. As a result of symmetry breaking, the ground state energy increases by 3.9 eV per unit-cell, which implies that the broken symmetry structure is less stable compared to the symmetric case. It should be noted that the effect of symmetry breaking on the ground state energy of the structure is significant since the energy difference (3.9 eV) is approximately two thirds of the cohesive energy of the pure structure (5.58 eV). To account for the effect of symmetry, the ground state energies of all possible cases leading to the same doping level were calculated separately and it was found that, at each doping level, the most favorable structure was the one that had the most symmetrical distribution of Na atoms in the interlayer region. In Fig.~\ref{fig2}b, we present the most favorable distributions of alkali atoms within the unit-cell for each doping level. These results imply that, after alkali activation, the alkali atoms in the C-S-H structure are going to evolve towards a symmetrical distribution in the interlayer region provided that sufficient time is allowed for the structure to settle into its ground state. This is likely controlled by the diffusion barriers of Na and Ca atoms in the interlayer.

\subsection*{Migration of interlayer atoms}
We calculated the diffusion barriers of the interlayer Na and Ca atoms in our clinotobermorite Na/Ca = 0.25 model structure by analyzing the energy landscapes of these atoms. The energy landscape of Na  presented in Fig.~\ref{fig3} was obtained by manually placing a substituted Na atom at various positions in the interlayer region and then recalculating the total energy for each position of the Na. The same procedure was repeated for a interlayer Ca.  During the optimization process, the Na (or Ca) atom was kept fixed at a particular position on the horizontal \textit{ab} plane, but its \textit{c} coordinate was allowed to move. This enabled us to scan a 3D energy path for the interlayer atom and project it onto the 2D plane. For this calculation, we placed the Na (or Ca) atom at 100 different points in the interlayer region that were uniformly distributed on the \textit{ab} plane. After calculating total energies for the 100 different locations of the Na (or Ca) atom, the energies corresponding to the remaining locations were determined by interpolation of the calculated data. Our results indicate that the Na atom needs to overcome a barrier of 1.6 eV to migrate in the interlayer region whereas this value is 1.8 eV for the Ca atom. However, both of these values are much lower than the cohesive energy of clinotobermorite (5.58 eV) which implies that Na and Ca can move within the interlayer region without breaking the integrity of the structure. 

The ability for Na and Ca to diffuse in the interlayer region of C-S-H will dictate whether a symmetric distribution of Na can be attained if the C-S-H phase has already precipitated. This information is extremely difficult to determine experimentally, and therefore molecular dynamics simulations are often used to calculate the diffusion coefficients and associated activation barriers (i.e., energies). Qomi et al.\cite{qomi2014combinatorial} reported energy barriers for the self-diffusion of water molecules in the interlayer spacing of C-S-H from their force-field molecular dynamics simulations of 150 different structures with varying Ca/Si ratios. They found that the barrier ranged from 0.08 to 0.12 eV, which is an order of magnitude lower compared with the Na (and Ca) energy barrier reported above. Although the energy barrier for interlayer Ca in C-S-H could not be found in the literature, quantitative data on alkali (and water) diffusion in related layered minerals is readily available. For Na-montmorillonite (2:1 layered aluminosilicate clay mineral), the Na and water molecule energy barriers were found to be roughly the same, falling between ~0.15-0.20 eV.\cite{greathouse2016molecular} Hence, it is significantly more difficult for Na (and Ca) to rearrange in the interlayer region of our model C-S-H system (clinotobermorite) compared with existing simulation literature on C-S-H. This may be due to the presence of cross-linking between the silicate chains in clinotobermorite, which is absent in pure C-S-H gels but arises as alkalis (and aluminum) are incorporated into the gel.\cite{myers2013generalized} Therefore, the ability for the gel to have an even (symmetrical) distribution of Na throughout its structure will likely be controlled by the initial formation mechanisms, where slower formation kinetics would facilitate the development of a more thermodynamically stable gel.

\subsection*{Effect of the alkali concentration on stability}
Having shown the migration behavior of the interlayer atoms, we subsequently evaluated the thermodynamic stabilities of the alkali doped clinotobermorite structures for different degrees of Na doping. For this purpose, we calculated the formation energy of the Na doped structure with respect to the pure structure, which is given as the difference between their cohesive energies, defined by $E_{f}[doped] = E_{c}[doped] - E_{c}[pure]$. Using this notation, the formation energy of the pure case is set to zero, negative values of $E_{f}$ imply less favorable structures whereas positive values of $E_{f}$ indicate more favorable structures as compared to the pure case. Fig.~\ref{fig4} shows the variation of formation energy with respect to the doping level, where it is clear that mild levels of Na doping result in structures with comparable stability to the pure case. For the Na/Ca = 0.05 and 0.08 doping ratios, structures with symmetrical distribution of Na atoms are only 50 meV less favorable than the pure clinotobermorite structure. However, if we consider the case with asymmetrically distributed Na atoms, the stabilities are significantly reduced despite low levels of doping. For example, for clinotobermorite with a doping ratio of Na/Ca = 0.08, the structure with asymmetric distribution of Na atoms is less favorable than the pure case by 3.90 eV. Similarly, the energy difference between the symmetric and asymmetric distributions is 4.16 eV at Na/Ca = 0.22. This behavior is prevalent for all doping levels as indicated by the difference between the red and black curves in Fig.~\ref{fig4}. Hence, different structures with the same level of alkali doping can have significantly different thermodynamic stabilities, even for very low doping levels.

As the doping level increases above Na/Ca = 0.08, the formation energy begins to decrease significantly even for the most symmetric arrangements, however this drop in energy plateaus when the Na/Ca doping ratio reaches 0.21. Above this level of doping, the formation energy resaturates at a new equilibrium value. This behavior suggests that there are two parameters affecting the overall stability of the alkali doped C-S-H, (i) level of doping and (ii) the symmetry of the alkali atoms within the structure. For low doping values, the symmetry of the alkali atoms within the interlayer region is the dominant factor. By ordering the Na atoms in a symmetric pattern, it is possible to obtain alkali doped structures which are similar in stability to the pure case. On the other hand, although symmetry still enhances stability for high doping values, the alkali doped structure is less favorable than the pure case.  Given that the Na/Ca ratios for alkali-containing C-S-H gels studied experimentally exceed 0.22, there is a high likelihood that the sodium-containing aluminum-substituted C-S-H (denoted as C-(N)-A-S-H) gel formed in Ca-rich AAMs and blended OPC cements may be less thermodynamically stable compared to the pure C-S-H gel case. However, additional research is required to assess the impact of aluminum content, finite silica chain length and nanoscale disorder on the formation energies and associated phase stability.

\begin{figure}
\includegraphics[width=8cm]{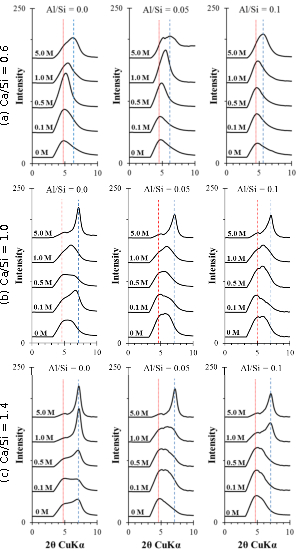}
\caption{XRD patterns of C-(N)-A-S-H gels with a Ca/Si ratio of (a) 0.6, (b) 1.0, and (c) 1.4. The red dotted line at $2\theta = 4.9$ defines the region (0.0\degree~to 4.9\degree) that may contain artifacts from the beamstop. Any peak outside this region is therefore considered real. The dashed blue line shows the position of the maximum intensity peak in each set. NaOH concentrations (in moles, M) in which the samples were synthesized are shown on the curves.}
\label{fig6}
\end{figure}

\begin{figure*}
\includegraphics[width=15cm]{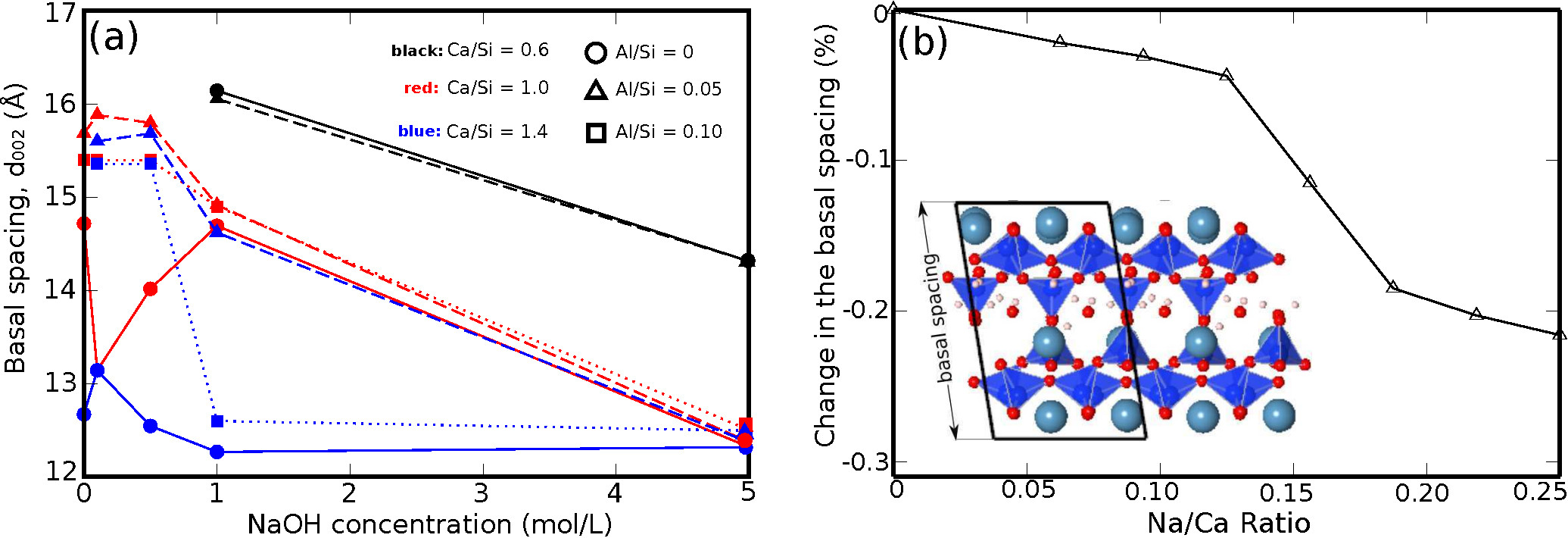}
\caption{(a) Experimentally-determined basal spacing of C-(N)-A-S-H gels as a function of NaOH concentration of the solution used for their synthesis. Circles represent Al/Si = 0.0, triangles represent Al/Si = 0.05, and squares represent Al/Si = 0.1.  Black, red and blue lines represent Ca/Si=0.6, Ca/Si=1.0 and Ca/Si=1.4, respectively. The interlayer spacing is based on the center of mass of the 002 Bragg peak, obtained by considering the intensities between three and nine degrees $2\theta$ if assuming Cu-K$\alpha$ radiation. Due to interference from the beam-stop, some samples did not display a distinct 002 peak and hence those have been omitted from the above plots. (b) Computationally-determined variation of the basal spacing of alkali doped clinotobermorite crystal with Na content in the interlayer region.}
\label{fig7}
\end{figure*}

To evaluate the effect of the Na/Ca ratio on the destabilization of clinotobermorite, we have generated electron density contour plots between the silicate tetrahedrons in the interlayer region for two representative Na doping levels (Na/Ca = 0.05 and 0.25), as shown in Fig.~\ref{fig5}. For both calculations, the isocontour interval was chosen to be the same (0.05 eV/ \AA $^2$) where a steeper gradient indicates denser electron accumulation and hence stronger chemical bonding. As seen in Fig.~\ref{fig5}, the silicate tetrahedrons are bound more strongly to each other in clinotobermorite for the lower Na/Ca ratio. This can be attributed to the geometrical alterations taking place in the interlayer of clinotobermorite as we increased the Na/Ca ratio, potentially making the higher Na/Ca ratio structure more susceptible to degradation when exposed to aggressive environments (such as acidic molecules).

\subsection*{Basal spacing}
Alkali doping is also seen to affect the basal spacing of the pure structure, which is one of the experimentally observable properties for characterizing C-S-H. The basal spacing is defined as the distance between consecutive Ca layers in the direction of the \textit{c} axis as shown in Fig.~\ref{fig1}b. Experimentally this distance is typically measured from the \textit{002} Bragg reflection in X-ray/neutron diffraction data for C-S-H and related model crystalline phases. From XRD patterns shown in Fig.~\ref{fig6}, basal spacings have been extracted using Bragg's law and plotted as a function of NaOH concentration in Fig.~\ref{fig7}a. It can be seen that gels with higher Ca/Si ratios tend to have smaller basal spacings for all alkali levels.  Moreover, it is clear that, increasing the alkali concentration tends to lead a decrease in the interlayer spacing for C-(N)-A-S-H gels for all Ca/Si and Al/Si ratios. Such a decrease in basal spacing has been reported by previous researchers,\cite{bach2013retention, l2016alkali} however a clear explanation for this reduction in basal spacing has been missing. While we do not discount the large role of water in governing the magnitude of basal spacing for such layered systems, especially C-S-H gels which undergo critical rearrangements upon drying,\cite{yang2018drying} based on our simulation of the basal spacing as a function of Na/Ca ratio, as shown in Fig.~\ref{fig7}b, we hypothesize that the geometrical positions of Na atoms in the interlayer also play role in the variation of basal spacing.

According to our simulations on the clinotobermorite crystal,  the basal spacing decreases gradually as Na replaces Ca atoms in the interlayer region and at the same time the structure expands along the horizontal \textit{ab} plane. When the Na/Ca ratio exceeds 0.11, the basal spacing starts to decrease steeply.  However, once the Na concentration increases beyond Na/Ca = 0.20, the decrease in the basal spacing slows down which indicates that it is no longer favorable for the structure to shrink. During this process whereby the basal spacing decreases, the Na atoms can migrate within the interlayer as expected due to their lower diffusion barrier compared with Ca atoms, as presented in Fig.~\ref{fig3}. Thus, the effect of Na concentration on the geometry of the unit-cell has three components: (i) the structure shrinks in the vertical direction due to the smaller atomic radius of Na with respect to Ca, (ii) Na can migrate easier than Ca in the interlayer region due to its lower diffusion barrier, and (iii) as the Na atoms migrate away from each other the structure further shrinks in the vertical direction.   These geometrical reconstructions distort the the bond angles/distances between the atoms in the interlayer region. Eventually the bond strength of the cross-linked silicate tetrahedrons decreases (Fig.~\ref{fig5}). We also note that, the variation of basal spacing with doping level presented in Fig.~\ref{fig6} shows a similar trend to the variation of formation energy presented in Fig.~\ref{fig4}.

\subsection*{Conclusion}
In conclusion, using first principles calculations we show that the thermodynamic stability of alkali doped C-S-H (using clinotobermorite as the model system) is closely related to the geometrical arrangement of the alkali atoms within the structure. The results reveal that as the Na atoms become more evenly distributed and attain a symmetrical pattern, the overall stability of the structure significantly increases. Our calculations show that the alkali atoms can migrate within the structure, but the energy barrier for diffusion is an order of magnitude larger than that for non-cross-linked C-S-H gel and related layered minerals reported in the literature. Hence, it is possible to obtain the same alkali concentration in C-S-H with several different arrangements of the alkali atoms within the interlayer. However, the structures with same doping levels can have significantly different degrees of stability depending on the way that the alkali atoms are distributed in the interlayer region, and we show that the thermodynamic stability of the structure increases as the alkali atoms are distributed more symmetrically. Depending on the doping level and symmetry of alkali atoms it is possible to have structures with comparable stabilities to the pure C-S-H, however the stability of the alkali doped structure decreases significantly when the Na/Ca ratio exceeds 0.11. Our calculations revealed for the first time that Na doping directly tends to a decrease in basal spacing, in agreement with experimentally obtained XRD data. The interplay between the nanoscale symmetry and the stability of the alkali doped structure observed here for clinotobermorite (model C-S-H structure) paves the way for future investigations concentrating on the effects of finite silicate chain length and aluminum.
\subsection*{Methods}
Our predictions have been obtained from state of the art first-principles pseudopotential calculations based on the spin-polarized DFT within generalized gradient approximation (GGA) including van der Waals corrections.\cite{grimme2006semi} We used projector-augmented wave potentials (PAW)\cite{blochl94} and the exchange-correlation potential has been approximated with Perdew-Burke-Ernzerhof (PBE) functional.\cite{pbe} The Brillouin zone (BZ) was sampled in the Monkhorst-Pack scheme, where the convergence in energy as a function of the number of \textbf{k}-points was tested. The \textbf{k}-point sampling of (3$\times$3$\times$3) was found to be suitable for the BZ corresponding to the primitive unit-cell of 11{\AA} clinotobermorite. For larger supercells this sampling was scaled accordingly. Atomic positions were optimized using the conjugate gradient method, where the total energy and atomic forces were minimized. The energy convergence value between two consecutive steps was chosen as $10^{-5}$ eV. A maximum force of 0.1 eV/\AA~ was allowed on each atom. Numerical calculations were carried out using the VASP software.\cite{kresse1999ultrasoft, vasp}

SEM measurements were performed on a Quanta 200 FEG Environmental-SEM under a water vapor environment. For the XRD analysis, a total of 45 different C-(N)-A-S-H gels were synthesized using Ca(OH)$_2$, SiO$_2$, and NaAlO$_2$ in an inert N$_2$ environment by varying the Ca/Si (0.6, 1.0, 1.4) and Al/Si (0, 0.05, 0.1) ratios, and using NaOH solutions of varying concentrations (0.0, 0.1, 0.5, 1.0, and 5.0 M). The water/solid weight ratio was 45 and the synthesis duration was 1 month after which excess of filtrate was removed via vacuum filtration. The precipitated gel was then dried overnight in a dry N$_2$ filled glovebox. Afterwards, the powder samples were loaded into polyimide capillaries for X-ray measurements. X-ray diffraction experiments were conducted on the 11-ID-B beamline at the Advanced Photon Source, Argonne National Laboratory, under ambient temperature conditions. A Perkin-Elmer amorphous silicon two dimensional (2D) image plate detector was used and each capillary was analyzed at an X-ray wavelength of 0.2115 \AA. 2D to 1D data conversion was carried out using the program Fit2D with CeO$_2$ as the calibration material.\cite{hammersley2016fit2d}

\subsection*{Acknowledgments}
We acknowledge funding from the Wilke 1989 Innovation Fund (Princeton University) and the Princeton Center for Complex Materials, a MRSEC supported by NSF Grant DMR 1420541. The calculations presented in this article were performed on computational resources supported by the Princeton Institute for Computational Science and Engineering (PICSciE) and the Office of Information Technology's High Performance Computing Center and Visualization Laboratory at Princeton University. We acknowledge the personnel at the 11-ID-B beam line which is located at the Advanced Photon Source, an Office of Science User Facility operated for the U.S. DOE Office of Science by Argonne National Laboratory, under U.S. DOE Contract DE-AC02-06CH11357. 

\bibliography{arxiv.bbl}

\end{document}